\documentclass[aps,prl,preprint,superscriptaddress,floatfix]{revtex4-1}

\usepackage{graphicx}
\usepackage{graphicx,epstopdf}

\begin{document}

%\preprint{}

\title{A Simple Pendulum Determination of the Gravitational Constant}

\author{Harold V. Parks}
\email[]{hvparks@sandia.gov}
\affiliation{JILA,  University of Colorado and National Institute of Standards and Technology, Boulder, CO 80309, USA}
\affiliation{Sandia National Laboratories, Albuquerque, NM 87185, USA}

\author{James E. Faller}
\affiliation{JILA,  University of Colorado and National Institute of Standards and Technology, Boulder, CO 80309, USA}

\date{\today}

\begin{abstract}
We determined the Newtonian Constant of Gravitation $G$ by interferometrically measuring the change in spacing between two free-hanging pendulum masses caused by the gravitational field from large tungsten source masses.  We find a value for $G$ of $(6.672\:34 \pm 0.000\:14) \times 10^{-11} \:\mathrm{m}^3 \,\mathrm{kg}^{-1} \,\mathrm{s}^{-2}$.  This value is in good agreement with the 1986 Committee on Data for Science and Technology (CODATA) value of $(6.672\:59 \pm 0.000\:85) \times 10^{-11} \:\mathrm{m}^3 \,\mathrm{kg}^{-1} \,\mathrm{s}^{-2}$ [Rev. Mod. Phys. {\bf 59}, 1121 (1987)] but differs from some more recent determinations as well as the latest CODATA recommendation of $(6.674\:28 \pm 0.000\:67) \times 10^{-11} \:\mathrm{m}^3 \,\mathrm{kg}^{-1} \,\mathrm{s}^{-2}$ [Rev. Mod. Phys. {\bf 80}, 633 (2008)].
\end{abstract}


\maketitle

%\section{}
Measurements of the gravitational constant $G$ have a very long history, that dates back to the birth of modern experimental science.  This precision measurement requires that the weak gravitational pull of a well-characterized source mass be measured to a high accuracy.  It is a supreme test of an experimental physicist to cleanly pull this signal out of the inevitable sea of perturbing influences.

Traditionally, $G$ is measured with a torsion balance.  In 1798, Cavendish and Michell reported numbers from a torsion balance that could be used to calculate $G$ to within about 1\% of its true value \cite{cavandish}.  It took nearly 200 years to improve on this accuracy by two orders of magnitude; in 1982 Luther and Towler reported a value of $G$ with an uncertainty of slightly less than 1 part in $10^4$ from a torsion balance experiment \cite{LutherAndTowler}.  This measurement became the principal basis of the accepted value of $G$ (Committee on Data for Science and Technology, CODATA 1986 \cite{codata86}) for over a decade.  

However in 1995, Kuroda pointed out that anelasticity in a torsion fiber (a frequency dependence of the restoring force due to material properties of the fiber) had the potential to cause a significant error at the level of uncertainty quoted by Luther and Towler \cite{kuroda}.  A number of new determinations of $G$ followed.  Many of these used a torsion balance in a mode that minimized the effects of the fiber anelasicity \cite{uwash,msl,bipm,hust,LANL,karagioz}, while several used other methods such as replacing the torsion balance with a simple pendulum \cite{wup99,*wup02} or a beam balance \cite{zurich}.  The lowest reported uncertainties from this new slate of measurements approach 1 part in $10^5$ \cite{uwash,hust,zurich}.   The CODATA recommended $G$ value has now shifted by 2.5 parts in $10^4$ from the Luther and Towler number, though the CODATA uncertainty remains at 1 part in $10^4$ because of some conflicting results \cite{codata06}.

Our determination uses a simple pendulum method similar to that of Kleinevo{\ss} et al.\ \cite{wup99,*wup02}.  By using a laser rather than a microwave interferometer and by better controlling the mass geometries, we have achieved a standard ($1\sigma$) uncertainty of 2.1 parts in $10^5$ for our value of $G$, which is an order of magnitude lower than the Kleinevo{\ss}  et al.\ result.  This uncertainty is within a factor of $\sqrt{2}$ of the lowest uncertainty $G$ value reported to date \cite{uwash}, but differs from this number by over $10\sigma$!  (We are  $2.9\sigma$ below the current CODATA value \cite{codata06} because of its larger uncertainty.)  We base our value on data taken in 2004, and in the interim, we have been unable to find a likely source for this discrepancy.  So, having checked and rechecked our work, we must finally report our value as we have found it.  It lies within the $1\sigma$ uncertainty band of the original Luther and Towler number.

\begin{figure}%[b]
$\begin{array}{c}
\includegraphics[width=3.2in]{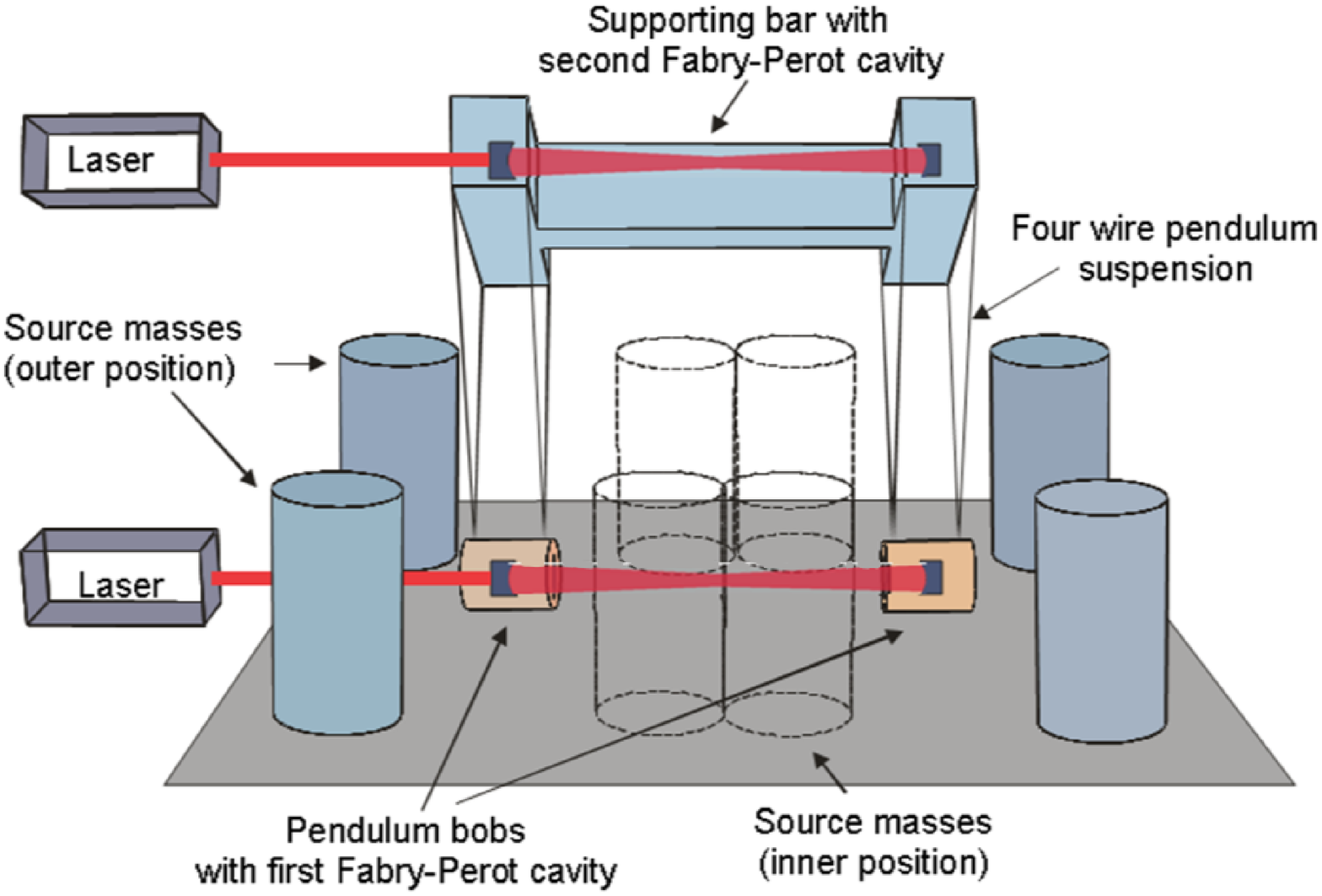} \\
\includegraphics[width=3.2in]{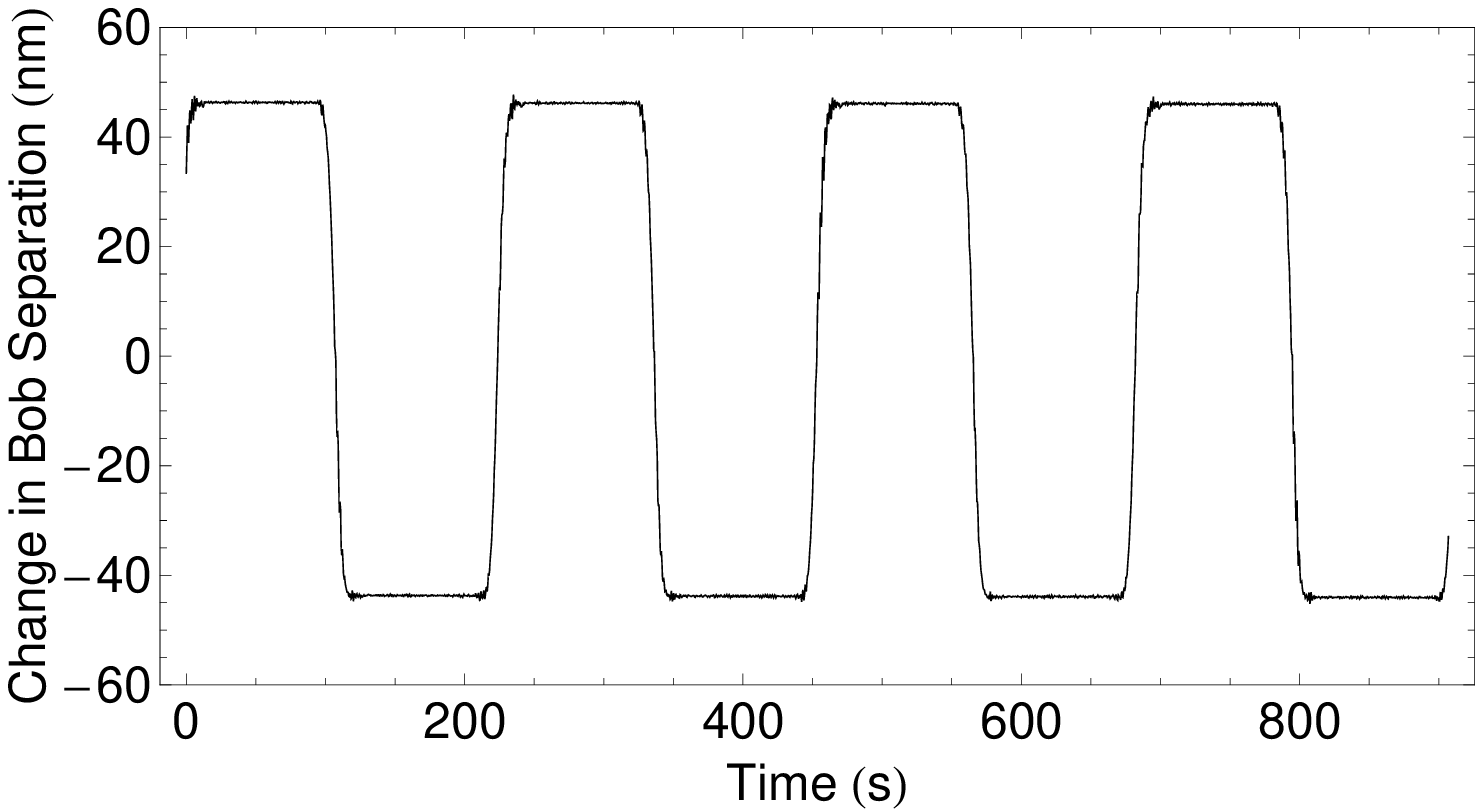}
\end{array}$
\caption{\label{schematic}
A schematic of the apparatus is shown on top.  A Fabry-Perot interferometer measures the spacing between the two pendulum bobs with respect to a suspension-point-located reference cavity.  The bobs are made of oxygen-free copper and have a mass of 780 g.  The pendulum length is 72 cm, and the spacing between the bob centers is 34 cm.  When the four 120 kg tungsten source masses (which are floated on air bearings) are moved from one position to another, the horizontal gravitational force on each pendulum bob changes by 480 nN, giving rise to a change in pendulum bob separation. Not pictured is the vacuum chamber that encloses the pendulums but not the source masses. Magnets (not shown) outside of the vacuum system and below the pendulum bobs damp the swinging motion of the pendulums so that the static deflection due to the gravitational pull of the source masses can be measured.  The gravitational signal is plotted on the bottom as the source masses are moved between the inner and outer positions several times (with the source masses pausing at each position for 80 s).  The 125 MHz change in the beat frequency between the laser locked to the pendulum cavity and the laser locked to the reference cavity corresponds to a 90 nm change in the pendulum bobs' separation.
}
\end{figure}

A schematic of our apparatus is shown in Fig.\ \ref{schematic}.  We find $G$ by balancing the gravitational pull of tungsten source masses against the restoring force of a simple pendulum:  
\begin{equation}
G\int\frac{\hat{\mathbf{z}}\cdot(\mathbf{x}-\mathbf{x}')\rho_s(\mathbf{x})\rho_t(\mathbf{x}')}{|\mathbf{x}-\mathbf{x}'|^3}\,d^3x\,d^3x' = -kz,
\end{equation}
where the source mass distribution $\rho_s$ corrected for displaced air and the test (pendulum bob) mass distribution $\rho_t$ are known.  The pendulum spring constant, up to some small corrections, is given by $k=m\omega^2$ with $m$ the bob mass and $\omega$ the angular frequency of the pendulum when it is set into free oscillation in a separate experiment. The four-wire pendulum design causes the bob to translate with very little rotation and constrains the bob to move only along the $\hat{\mathbf{z}}$ axis.  Since most (99.87\%) of the pendulum restoring force is from earth's gravity rather than from the material properties of a fiber, the pendulums behave very much like perfect springs for small displacements.  These springs are stiff compared to a torsion fiber, but this stiffness is offset by the ability of the laser interferometer to measure very accurately the change in the distance between the two pendulum bobs that occurs when the source masses are moved from one position to another.  

The most difficult aspect of any precision measurement experiment is understanding and controlling the major sources of uncertainty.  Though conceptually the experiment is very simple, Nature's cunning is in the details.  We sketch out the uncertainty sources here but a longer follow-up paper is planned to more fully describe the experimental details.   The uncertainties are summarized in Table \ref{budget} and are dominated by components related to the mass distributions.

\begin{table}
\caption{\label{budget}
The major components of uncertainty are listed here expressed in terms of each contribution to $\delta G/G$ in parts in $10^5$ at the $1\sigma$ level. The uncertainties in this table, along with all other uncertainties in this paper, are expressed as standard ($1\sigma$) uncertainties.
}
\begin{ruledtabular}
\begin{tabular}{l r}
Uncertainty Component&$\delta G/G (\times10^{-5})$\\
\hline
Six critical dimensions&1.4\\
All other dimensions&0.8\\
Source mass density inhomogeneities&0.8\\
Pendulum spring constants&0.7\\
Total mass measurement&0.6\\
Interferometer&0.6\\
Tilt due to source mass motion&0.1\\
Day-to-day scatter&0.4\\
\hline
Combined uncertainty&2.1\\
\end{tabular}
\end{ruledtabular}
\end{table}

The source masses are arranged so that, in both measuring positions, the pendulum bobs are at a saddle point in the gravitational field from the source masses.  This makes the gravitational signal quite insensitive to the position of the pendulum bobs relative to the source masses, though the signal does depend critically on the distance - perpendicular to the interferometer axis - between the two opposite pairs of source mass cylinders as well as the along-axis distance between the two adjacent source masses when they are in the inner position.  This geometry reduces the hardest part of defining the three-dimensional mass distribution to just six one-dimensional measurements.  We constructed a large caliper with a movable stand that could reach around the apparatus.  With this and a smaller caliper, we were able to measure the six critical separations with an uncertainty of about 3 $\mu$m. This measurement contributes a relative uncertainty of 1.4 parts in $10^5$ to our combined uncertainty.  The gravitational signal is much less sensitive to uncertainties in all the other dimensional measurements, but we also invested less effort in making these other measurements, which contribute a total of 0.8 parts in $10^5$ to the uncertainty budget.

Density variations within the source masses are also a significant contributor to the uncertainty of our final value.  The masses are made of an alloy of 95.5\% tungsten sintered with copper and nickel.  Because the cylinders were cast on their sides, our finding a density variation of 1 to 2 parts in $10^3$ across their diameters is not surprising. This density variation was measured by allowing individual billets to rotate freely in an air-bearing as well as by cutting apart one of the billets after the experiment was concluded.  The orientation of each source mass stack (as well as the orientation of the three billets that comprise it) was adjusted to cancel out, by as much as possible, the effect of this gradient on the total gravity signal.   We also rotated the stacks by $180\,^{\circ}$ halfway through the experiment to average out the effect of any residual linear component of the density gradient.  Based on the air bearing data, the expected fractional change in the gravity signal was $(2.4 \pm 0.5) \times 10^{-5}$ when the masses were rotated $180\,^{\circ}$.  We actually observed a fractional change of $(1.3 \pm 0.7) \times 10^{-5}$, in reasonable agreement with the calculated value.  The residual nonlinear density variations contribute an uncertainty of 0.8 parts in $10^5$ to the final result.

The total mass of the source mass configuration contributes 0.6 parts in $10^5$ to the uncertainty budget which includes both the uncertainty of the balance used to weigh the masses and the uncertainty in the density of the displaced air. 

The spring constant of each pendulum is obtained by setting the pendulums swinging (with the damping magnets removed) and recording the period of oscillation.  We make three corrections to the simple pendulum model.  The first correction is from the small, but non-zero, rotational inertia of the wires and amounts to a relative correction of $(7.5 \pm 0.1) \times 10^{-5}$ to the spring constants.  Second, we take into account the fact that the bobs rotate slightly as they translate.  This rotation occurs because the relative loading on the wires changes as the bob is displaced, causing the wires to stretch differentially.  This rotation results in a correction to the spring constants of $(5.8 \pm 0.4) \times 10^{-5}$.  Finally, we account for the force on the pendulum bobs from the damping magnets due to the diamagnetism of the bobs.  The horizontal force gradient was measured by translating the magnets and observing the resulting displacement of the pendulum bobs.  As copper is diamagnetic, the bobs were observed to move in the opposite direction from the magnets (confirming that there was no ferromagnetic contamination on or in the bobs).  

Because they are diamagnetic, there is also a small upward magnetic force on the bobs that reduces the effective value for $g$ on the bobs.  This force was evaluated by weighing the bobs with and without the magnetic field.  We find a total spring constant correction due to magnetic effects of $(-7.54 \pm 0.03) \times 10^{-5}$ for one pendulum and $(-7.34 \pm 0.01) \times 10^{-5}$ for the other.  

Corrections due to the finite amplitude of the swing during the pendulum frequency measurements and corrections due to the finite Q of the pendulums (with the damping magnets removed) are less than about 1 part in $10^6$ and were ignored.  The remainder of the uncertainty in the pendulum spring constants comes from scatter in the data used to measure the periods (0.5 parts in $10^5$) and the measured anelasticity of the pendulum wires (0.2 parts in $10^5$).

The pendulum bobs are slightly magnetized by the field of the damping magnets, and this makes them more sensitive to magnetic gradients than they otherwise would be.  (The field in the vicinity of the pendulum bobs is on the order of 0.01 T, and the susceptibility of the copper bobs is $-1 \times 10^{-5}$.)   However, residual fields from the damping magnets are on the order of a few hundred $\mu$T in the vicinity of the source masses and are too small by more than an order of magnitude to induce sufficient magnetization in the source masses to influence the bob position as the tungsten alloy used for the source masses has a susceptibility of $(6.6 \pm 0.3) \times 10^{-4}$.   Care was also taken to eliminate any error due to magnetic fields from the source mass drive motor.

Because we must move large source masses to generate the gravitational signal, care must be taken to reduce possible errors due to the change in mass loading on the apparatus.  The vacuum chamber that contains the pendulums straddles, without touching, the plate upon which the source masses ride.  Finally, the source mass support plate rests kinematically on the floor independently from the rest of the apparatus.  The center of mass of the 480 kg source mass configuration shifts by 0.2 mm when it is moved from the inner to the outer position because of a slight deviation from the planned values of the mass stop locations.  Though this is a small shift, the resulting change in floor tilt translates to a change in the pendulum bob separation because the pendulums differ in length by 0.3 mm.  We evaluated this effect by deliberately shifting the center of mass position of the source masses by a large amount and observing the effect on the pendulums (after removing the calculated gravitational signal).  Based on this data, we find a correction of $(-0.4 \pm 0.1) \times 10^{-5}$ to our $G$ value.

The compressed air that is fed to the air pucks under the source masses cools as it is released. Care was taken to ensure that the resulting thermal gradients did not cause an error in the final results.  A vacuum pump connected to a groove around the outer perimeter of the puck sweeps up the cool air before it escapes from under the puck.  Temperature measurements of different parts of the apparatus (the source masses, source mass support plate, and pendulum vacuum chamber) indicated that all parts were at the same temperature to within $0.1 \,^{\circ}\mathrm{C}$.  As a check of the temperature sensitivity of our apparatus, we raised the temperature of the source masses from the ambient $22 \,^{\circ}\mathrm{C}$ to between 30 and $40 \,^{\circ}\mathrm{C}$.  With the source masses at this elevated temperature, the pendulum signal changed by a factor $(4 \pm 22) \times 10^{-5}$ after correcting for the mass position change due to the thermal expansion of the apparatus.  We conclude that temperature effects have a negligible contribution to our uncertainty budget (aside from a term that we have included in the uncertainty of the dimensional measurements).

The laser interferometer contributes to the uncertainty budget mainly through any misalignment of the optical axis with respect to the pendulum bob motion as well as scatter (due to pendulum motion) in the data used to  determine the free spectral range.  We use He-Ne lasers locked to the pendulum and reference cavities with a Pound-Drever-Hall scheme.  About 1 $\mu$W reaches each Fabry-Perot cavity, and each cavity has a finesse of 4000.  Optical effects, such as stray reflections from the various optical components as well as radiation pressure on the bob mirrors, are negligible sources of uncertainty.

\begin{figure}
\includegraphics[width=3.2in]{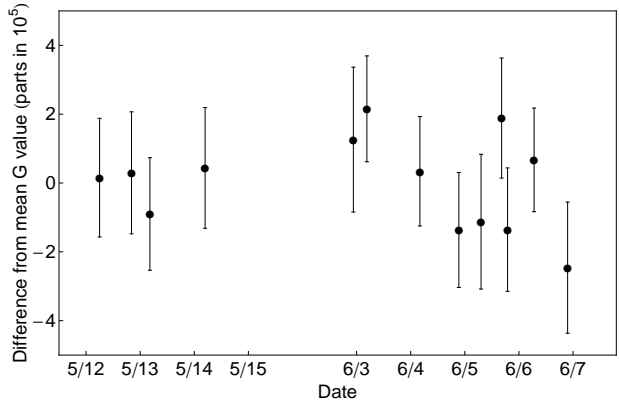}
\caption{\label{values}
The calculated $G$ values from data runs in May and June of 2004.  Each value is expressed as the fractional deviation $(\times10^{-5})$ from the mean value of $6.672\,34 \times 10^{-11} \:\mathrm{m}^3 \,\mathrm{kg}^{-1} \,\mathrm{s}^{-2}$.  The error bars include the uncertainty calculated from the scatter within each data set combined with the $1.5 \times10^{-5}$ relative uncertainty associated with the observed day-to-day variations in the source mass position.  The systematic components of the uncertainty, as listed in Table \ref{budget}, are not included in the error bars.
}
\end{figure}

A summary of the 13 data runs used in this determination of $G$ is shown in Fig.\ \ref{values} which gives the calculated G values from data runs in May and June of 2004.  Each run consists of between one and a half and seven hours of data like that shown in the bottom panel of Fig.\ \ref{schematic}.  During the time period covered in Fig.\ \ref{values}, the six critical source mass dimensions were measured eight times.  For each data point, the value of $G$ was calculated using the average of the source mass positions that were found before and after that run or that day's series of runs.  The source mass positions vary slightly from day-to-day because of movement of the stops as the 120 kg source masses are seated and variations in the force pressing the masses into the stops.  The standard deviation in these position measurements is 3.6 $\mu$m, which is expected to cause a standard deviation of $1.5 \times 10^{-5}$ in the signal from run to run.   This is very close to the standard deviation of $1.4 \times 10^{-5}$ actually seen in Fig.\ \ref{values}.  

During the gap between the 5/15 and 6/3 data, there was a large 50 $\mu$m shift in one mass position that occurred when the source masses slammed into the stops while we were trying to troubleshoot a faulty motor.  This collision caused a large shift in the raw signal, but no significant shift is seen in the $G$ values after the new positions were used in the calculations.  After the motor problem, the drive system required constant readjustment and three data sets were thrown out because the source masses were getting stuck before they were fully into the mass stops.  In addition, four of the data sets (the 5th, 9th, 11th, and 13th points in Fig. 2) were truncated after two hours when the signal became noticeably unstable towards the end of the run.

Between the data taken on 6/3 and 6/4 (the 6th and 7th data points), each source mass stack was rotated by $180\,^{\circ}$ to average out the linear density gradient across the source mass billets.  A correction based on the measured density gradient is included in the data shown in Fig.\ \ref{values}.  Nevertheless, the value for $G$ we give, calculated as the mean of the data before and after the $180\,^{\circ}$ rotation, does not depend on the value of this correction.

We have presented here our new determination of the Newtonian constant of gravitation. Great care was exercised in carrying out the experiment and in our detailed analysis. Having now completed our measurement, we are reminded of Cavendish's description of his 1798 experiment \cite{cavandish}: ``The apparatus is very simple.'' That statement also applies to the experiment that we report here.  We would add: ``The measurement is very hard.''

%\appendix
%\section{}

\begin{acknowledgments}
We thank Douglas S. Robertson for writing software to provide an independent check of our gravity field calculations as well as Hans Green, Blaine Horner, and Alan Patee for creating the apparatus.  We also thank Terry Quinn and Richard Davis for many helpful discussions.  H. Parks is grateful to the National Research Council for a NIST post-doctoral fellowship. Sandia National Laboratories is a multiprogram laboratory operated by Sandia Corporation, a wholly owned subsidiary of Lockheed Martin company, for the U.S. Department of Energy's National Nuclear Security Administration under contract DE-AC04-94AL85000.  This manuscript has been assigned report number SAND 2010-5164J by Sandia National Laboratories.
\end{acknowledgments}

\bibliography{jila_G}

\end{document}